# Pacific shallow lagoon high-resolution temperature observations

**Hans van Haren**

Royal Netherlands Institute for Sea Research (NIOZ) and Utrecht University, 1790 AB Den Burg, the Netherlands
hans.van.haren@nioz.nl

## ABSTRACT

The daily cycle of heating and cooling of the near-surface ocean may be quite different in a shallow lagoon with a few meters deep seafloor that can be heated directly by the sun. If important, the solar radiation will affect the local benthic communities. To study the physical processes associated with the daily cycle of south-Pacific lagoon Bora Bora, a vertical string of five high-resolution temperature sensors was moored at a 2-m deep site for 3 weeks. Besides the standard ocean warming (approximately during daytime) and cooling (approximately nighttime), the sensors show relatively highest temperature near the lagoon-floor during the warming phase and a weakly stable stratification towards the end of the cooling phase. During the warming phase, highly variable stratification is observed extending into the water column under calm weather and turbid waters, otherwise not. Under trade wind and clear waters, the lowest sensor(s) show(s) consistently higher temperature variability than sensors higher-up with spectral slopes indicative of shear- and/or convective turbulence. During the cooling phase, the lower sensor shows consistently very low variance (non-turbulent), while other sensors show a spectral slope around the buoyancy frequency evidencing weakly stratified waters supporting internal waves. These observations contrast with open-ocean near-surface observations of stable stratification during the warming phase and of turbulent free convection during the cooling phase. Thus, lagoons seem to more resemble the atmosphere than the ocean in daytime thermodynamics and possibly act as a natural solar pond with bottom conductive heating (when salinity compensates for unstable temperature variations).

# INTRODUCTION

The daily open-ocean near-surface heating response pattern relates with one crucial difference between the atmosphere and the ocean in that the atmosphere is cooled at a higher geopotential level than where it is heated, while the ocean is heated and cooled at its top (Sandström, 1908). As warm air/water is generally less dense than cold air/water, the daytime heating generates natural free vertical turbulent convective motions in the atmospheric boundary layer, resulting in an effective heat engine (Peixoto, and Oort, 1992). In contrast, solar radiation heats the ocean from above, with most radiation being absorbed near the surface resulting in stable density stratification and causing it to be merely a heat transporter with its heat engine being quite ineffective (Munk, and Wunsch, 1998). As a result, vertical (diapycnal) turbulent exchange is generally considerably weaker in the stratified ocean than in the atmosphere, the net radiation flux being positive. During the cooling phase, the near-surface ocean exhibits turbulent free convective overturning like the warming atmosphere, but for a limited near-surface range $O(10 \text{ m})$ due to the stable stratification that requires large mechanical turbulence to break down the stored potential energy (e.g., Brainerd, and Gregg, 1993; 1995; Price *et al*., 1986]. In shallow near-coastal waters and near the ocean surface in general, a relatively strong solar daily rhythm is thus expected not only in temperature at a given depth but also in temperature (density) stratification. The turbulent mixing associated with the turbulent convection due to cooling has values (Gregg *et al*., 1985) that are of the same order of magnitude as those associated with frictional flows in an energetic tidal channel (e.g., Korotenko *et al*., 2013) and with the breaking of large internal waves above steep topography in the deep ocean (e.g., van Haren et al., 2016).

In very shallow seas another physical process may be so dominant that it counteracts the above oceanic daily turbulence convection and stable stratification cycle. This process is the rather efficient (about 20% according to Hull, 1979) capturing of energy in so-called 'solar ponds': Near-bottom heating that may be stored in waters where a stable salinity-density stratification is found as observed in a Transylvanian lake (von Kalecsinky, 1902). As water is transparent to visible light but opaque to infrared radiation, a seafloor absorbs heat which can only escape into the water column via conduction and, especially when not stably stratified, higher-up via turbulent convection.

In this paper temperature data are presented that are reminiscent of the solar pond process. They are observed during periods of little rain and generally clear waters, so that solar insolation dominates temperature- and stratification-variations. In addition, the observations show internal waves supported by the stratification that dominate over turbulence activity during part of the cooling phase. The above physics process of daytime near-bottom heating may occur irrespective of, and add to, the 'biophysics' process of microenvironment heating the first millimeter above specific shallow water corals that has been demonstrated mainly in laboratory studies (Fabricius, 2006; Jimenez *et al*., 2008). A small-scale mooring holding five high-resolution temperature (T) sensors and a single low-resolution pressure (P), temperature, salinity (S) PTS-sensor (Fig. 1a) was deployed in about 2 m water depth near the anchor of a sailing vessel in the lagoon of Bora Bora, an island in the southern tropical Pacific (Fig. 1b). The aim of the set-up was to study details of the shallow lagoon's daily solar heating cycle and associated effect on temperature stratification at several depth levels. It was expected to observe alternating stably stratified and turbulent periods during the warming (approximately daytime) and cooling (approximately nighttime) phases, just like in open-ocean near-surface observations (e.g., Brainerd, and Gregg, 1993; Gregg *et al*., 1985; Price *et al*., 1986; Soloviev, and Lukas, 1997). Such open-ocean daily cycle has also been used to model the daily cycle in lagoons that have water-depths varying between a few and several tens of meters (e.g., McCabe *et al*., 2010). This model contrasts with recent modelling of a 0.1 m deep lake that is heated from below (de la Fuente, 2014). The present observations are aimed to resolve which process is dominant under what conditions in the parts of a typical Pacific lagoon that are a few meters deep.

**METHODS**

Eulerian temperature variations were observed using five stand-alone high-resolution NIOZ4 sensors moored at 2 m water depth with a relatively white sandy seafloor near the south-Pacific tropical atoll of Bora Bora at 16° 31.84′ S, 151° 42.27′ W (Fig. 1). According to Gischler (2011), the seafloor in the shallow less than 5 m deep back-reef sand apron on the eastern side of Bora Bora consists of skeletal grains and 30-50% of non-skeletal grains (cemented fecal pellets; aggregates). The mooring site is sheltered from surface waves and the ESE trade wind. The topography is bumpy, with vertical variations of up to 0.3 to 0.6 m that are 0.5 to 1 m apart. The mooring line was put in a small trough, in the vicinity of the anchor with its 1 kg weight buried. The total length of the mooring-line was about 1.8 m. Vertical sensor heights were measured at the 1 mm glass tip. They were at 0.1 m above the bottom (mab), 0.4, 0.7, 1.0 and 1.5 mab. The T-sensors sampled at a rate of 1 Hz. Between the uppermost two sensors a low-resolution Star-oddi P,T,S sensor was attached, which sampled at a rate of once per 600 s. Thus, at only one level, around 1.3 mab, T and S information was available having accuracy/resolution of ±0.1/0.032°C and ±1/0.02 g/kg, respectively. This sensor was mainly used to roughly detect near-surface fresh water and pressure variations. The line was held vertically by two plastic bottle floats on top providing about 0.5 kg net buoyancy. Meteorological information was obtained from 'Meteoblue' (https://www.meteoblue.com/en/weather/forecast/archive/bora-bora_french-polynesia_4034582) whilst visual turbidity information was noted in a logbook.

The NIOZ4 T-sensors are an improved version of NIOZ3-temperature sensors with basically the same characteristics, including a precision better than $5\times10^{-4}$°C and a noise level of less than $1\times10^{-4}$°C (see van Haren *et al*., 2009 for details). In normal mooring configurations, the clocks of all T-sensors are synchronized to a standard clock every 4 hours, so that an entire profile is measured to within 0.02 s. However, the somewhat bulky synchronizer was left out of the present small mooring line. Instead, prior to deployment the entire mooring line was submersed rapidly in different ambient temperature. The fast (<0.25 s) response of the T-sensors allows for time shifts to be corrected

Although calibrated to within a precision of about $1\times10^{-4}$°C in the laboratory before and after the year of deployments, the drift of the NIOZ4 sensors is about $1\times10^{-3}$°C/month after aging. The normal post-processing procedure corrects for this drift by forcing an average temperature profile to a smooth statically stable profile. This works well under the criteria that the number of sensors is large, typically 100, and that temperature can be used as a tracer for density variations via a tight relationship. In the present case, neither of these two criteria are (always) met. During rainfall, the influx of fresh water can provide a too large contribution of salinity on density variations compared with temperature contributions. This is remedied here by focusing on a period with virtually no rain resulting in little observed variability in S near the surface and confirmed from meteorological information. The number of T-sensors cannot be altered however, and a subjective fitting for an arbitrary, weakly unstable nighttime period between days 216.8 and 217.2 is performed to correct for the drift. The small number of sensors also precludes estimates of turbulence using the method of overturning scales. Instead, qualitative turbulence information is obtained from scalar temperature variance spectra (Tennekes, and Lumley, 1972), focusing on the frequency range around N where a transition from stable stratification supporting internal wave propagation to unstable turbulence is expected. Internal waves are driven by any (shear) flow interacting with topography and have frequencies between inertial frequency f and N (e.g., LeBlond, and Mysak, 1978). Here the mean buoyancy frequency is defined as $N = (-g/\rho \cdot d\sigma_\theta/dz)^{1/2}$ using the local thermal expansion coefficient of $\Delta\sigma_\theta/\rho\Delta T \approx 3.1 \times 10^{-4} °C^{-1}$ for ambient P, T and S using GSW-software (IOC, SCOR, and IAPSO, 2010), with g denoting the acceleration of gravity, $\rho$ a reference density and $\sigma_\theta$ the potential density anomaly referenced to the surface. Variations in $\sigma_\theta$ are determined from temperature variations.

## OBSERVATIONS

Like near-surface open-ocean waters, lagoon waters generally warm up during the day and cool during the night (Fig. 2a). The daily rhythm is apparent in all our temperature records, although its magnitude varies between days and between sensor depths. By computing the temperature variations relative to that of the lowest (fifth) T-sensor (Fig. 2b), one observes positive and negative values of $\Delta T_{i5} = T_i - T_5$, i = 1,…,4, which vary over the range $-0.4 < \Delta T_{i5} < 0.7°C$. A positive relative temperature gives rise to a stable density stratification and a negative value suggests turbulent free convection (provided the salinity contributions to density variations are negligible). Sometimes, e.g. around days 215 and 230, and at some sensors positive relative temperature is observed during the warming phase as in general ocean heating. However, quite often the relative temperature is negative during the warming phase, with temperature at the lowest (or second-lowest) T-sensor being highest. Notably, during days 212 and 213, temperature at the lowest sensor is higher than that at all other sensors by up to 0.4°C. More in general, the relative temperature is -0.2±0.1°C and also temperature at the second-lowest sensor is higher (as is understood from less negative relative temperature in Fig. 2b) than at the sensors above. During the cooling phase, the relative temperature initially becomes negative shortly after sunset, as is visible in records of near-surface and near-bottom stratification, i.e. the vertical temperature difference between neighbouring sensors, in Fig. 2c. (Note the change in y-axis scale compared with Fig. 2b). However, the water at the upper T-sensors cools slower than at lower sensors, so that the temperature difference becomes positive over time. The positive peak in T-difference during the cooling phase is generally observed to occur at sunrise, e.g. around days 220 and 230. After sunrise, cooling continues for approximately another 1-3 hours (Fig. 2a) while the T-difference drops to near-zero (Fig. 2c).

In Fig. 3 the first five days of the record are presented in more detail, which comprise various dynamics. These five days were without precipitation, without detected fresh-water influx near the surface. Days 212, 213 showed significant easterly trade wind with speeds of about 10 m s$^{-1}$ (meteorological information) and agitation with clear waters over the mooring (logbook information). Days 214, 215 were relatively calm with wind speeds of less than 6 m s$^{-1}$ and with some turbid waters over the mooring (probably due to materials from the main island to the northwest). The vertical temperature variation on day 216 is representative for the next 10 days of the record. The salinity variation (Fig. 3c) does not show negative spikes (strong decreases in value), evidencing no freshwater (rainfall) influx. Salinity follows statistically insignificantly a daily periodicity that is not related with the predominant semidiurnal tidal pressure variation. The barotropic surface tidal variations have an amplitude in pressure of 0.1 dbar, or about ±0.1 m in surface height variation (Fig. 3b). The quasi-daily salinity variation most likely reflects indirect effects of fouling, with imprecise T- and conductivity (C-) data mismatches resulting in erroneous S-data. The P,T,S-sensor was not cleaned during its operation. The basic mean level of salinity steadily declined with time (hence the present mean value is 32.8 psu, whereas near the beginning of the record values exceeding 35 psu were measured). Much value can therefore not be extracted from these data. The T-S relationship could not be determined properly also because of the large noise of the P,T,S-sensor.

The maximum observed stratification $\Delta T > 0$ (Fig. 3d) yields a local buoyancy frequency of $N_{1.5m} \approx 4 \times 10^{-2}$ s$^{-1}$. This vertical density stratification supports freely propagating internal waves with periods as short as 150 s. Possibly, and as will be inferred from spectral observations below, higher frequency ($\sigma$) internal wave motions up to $\sigma = 10^{-1}$ s$^{-1}$ (supported by maximum small-scale $N_{max} \approx 1400$ cpd, short for 'cycles per day') exist briefly due to unresolved thin (<0.35 m) layer stratification. The typically observed 'low value' stratification during the cooling phase yields $N_l \approx 3 \times 10^{-3}$ s$^{-1}$ ($\approx 45$ cpd).

The largest stratification in Fig. 3d is found during the warming phase and near the surface, but only on calm days (meteorological information) like on days 214 and 215 with turbid waters (logbook information). On days 212 and 213 of trade wind, waters were clear and the warming phase showed little stratification except between the two uppermost T-sensors near the end of warming (Fig. 2c). Instead, near-bottom temperature increased most of all (Fig. 3a) and temperature 'instabilities' $\Delta T < 0$ were observed in Fig. 3d between the lowest two sensors 4 and 5 (day 212) and also sensors 3 and 4 (day 213).

The largest negative temperature differences are observed around mid-day, when the sun is highest. The larger T-increase at especially the lowest sensor, and to a lesser extent at the second lowest sensor, are accompanied by an increase in short-term T-variability. This starts about 2.5 h after sunrise and stops within ±0.5 h from sunset. It evidences the sun heating the bottom and thus the water from below. Unless compensated by salinity stratification, this would create a static instability throughout the warming phase. The rapid temperature variability observed at the lowest sensor(s) points at such, but more qualitative information is gained from spectral analysis. Spectra are also used to investigate whether shear-induced turbulence prevents convection from developing towards the surface, as the T-time series observations suggest.

During the warming phase, the apparent T-instability, especially near the bottom, is reflected in T-variance spectra (Fig. 4a). The spectra are smooth ensembles of 13 ~daytime periods of 10 h of data. The spectrum of the uppermost sensor shows lowest variance during the warming phase. This spectrum scales with a frequency rate of $\sigma^{-5/3}$ (which yields a slope of -5/3 on a log-log scale) between 20 and 2000 cpd before rolling off. The -5/3-slope reflects the inertial subrange of turbulence, which extends over a range of two orders of magnitude. The inertial subrange represents a passive scalar shear-induced turbulence (Tennekes, and Lumley, 1972). In contrast, the more energetic spectra from data of the lower sensors (here presented by the fifth sensor) show a slope between -5/3 and -1 in the range between 50 and 2000 cpd. This intermediate slope of about -1.3 reflects a dominant active scalar convective turbulence (Cimatoribus, and van Haren, 2015). (The slope of -1 is indicative for open-ocean internal waves, van Haren, and Gostiaux, 2009).

During the cooling phase, the 13 ~nighttime smooth spectra of 10 h of data (Fig. 4b) show different slopes. In comparison with the warming phase spectra, the T-variance is smaller at all frequencies and at all sensors. In the frequency range under investigation, the now smoothest temporal variations and smallest spectral variance are observed by the lowest sensor. The spectrum for this fifth sensor shows a steep slope between -2 ($\sigma < 5000$ cpd; indicative of finestructure contamination, Phillips, 1971) and -3, before ending in white noise at $\sigma > 2 \times 10^4$ cpd. (Note that in all spectra the entire vertical range of variance is ten orders of magnitude). The uppermost sensor (representing all other sensors) shows a slightly more convective slope between -1 (around N) and -5/3 (between 20 and 200 cpd). At frequencies $\sigma < 30$ cpd in the presented range, the slope equals -1 indicative of dominant internal waves (van Haren and Gostiaux, 2009), at all sensors.

Coherence spectra comparing data from neighbouring T-sensors (Fig. 4c,d) confirm the power spectra observations. Coherence is generally less during the warming phase (Fig. 4c) than during the cooling phase (Fig. 4d), but for σ < 100 cpd only. This suggests (highly coherent) internal wave motions to be relatively more important during the cooling phase, at all sensors. For σ > 100 cpd, the coherence during the warming phase is larger than that during the cooling phase, except near the seafloor for 100 < σ < 1000 cpd. The large extent of significantly coherent signals for σ > 2$N_{max}$ points at extended turbulence activity in non-homogeneous waters (van Haren *et al*., 2016).

Other moderately smoothed spectra of 10-h periods from days 214 (weak wind, turbid waters) in Fig. 5 and 212 (strong wind, clear waters) in Fig. 6 show similar features that are perhaps somewhat more pronounced. During the warming phase, the data from upper (and middle) T-sensors showed a -5/3 slope over most of the investigated range for σ < 2000 cpd, while the more energetic lowest sensor data showed a slightly less steep slope between -5/3 and -1 for σ < 400 cpd (Fig. 6a). During the cooling phase, the small variance at the lowest sensor slopes steeply at rates <-2 down to -3. Spectra from data of the other sensors also showed less variance than during the warming phase, but had an average slope of -5/3 between 20 and 2000 cpd, with distinct patches of slopes of -1, especially around N, evidencing high-frequency internal wave activity.

While the coherence spectra from the cooling phase were approximately similar for data from all sensors during both periods, dropping to insignificance levels between 100 < σ < 1000 cpd, coherence was completely different between the two periods for the warming phase. During the warming phase, correspondence was insignificant (non-coherent) between all depth levels at all frequencies for days 214, 215 (Fig. 5c), while being coherent between all levels, except near the seafloor, at nearly all σ < $10^4$ cpd for days 212, 213 (Fig. 6c).

## DISCUSSION

It seems that shallow lagoons can resemble the atmosphere by being heated from below. Part of the solar radiation penetrates to the seafloor when waters are clear. If so, either its partial reflection or conduction heat the water from below. As a result, near-bottom temperature and temperature variance (of fluctuations) are high during the warming phase. The T-fluctuations partially suggest strong turbulent motions. The observed abrupt change from strongest to weakest turbulence near the seafloor relative to higher-up at sunset (and vice versa two hours after sunrise) further demonstrates the directly solar-driven heating from below. In the absence of (knowledge on) salinity contributions to density, one would consider turbulent motions due to heating from below to be primarily convectively driven.

However, the qualitative analysis of scalar temperature variance spectra focusing on the frequency range around N shows that the created turbulence has a dominating character of shearing motions, especially away from the bottom. This suggests that shear-induced turbulence prevents convection from developing further upwards towards the surface, or that vertical motions related with convection generate secondary shear (Li, and Li, 2006), or, perhaps unlikely, that horizontal motions are generated following differential heating over slightly different water depths due to the bumpy topography with possibly varying reflective/absorption properties.

Although near-bottom salinity observations were lacking, little evidence was found for an effective solar pond in the Bora Bora lagoon that retains large heat content close to the bottom via turbulence-dampening salt stratification. During the cooling phase, in fact at sunset, all near-bottom surplus heat content was lost rapidly instead of being maintained throughout the night. This indirectly evidences weak salinity stratification. The observed minimum vertical temperature difference was $\Delta T \geq -0.4°C$ and spectra did not show clear internal wave slopes evidencing salinity/density stratification during the warming phase. The weak cooling phase stratification, generated by $\Delta T \approx +0.01°C$ providing $N_l$, is roughly supported by internal wave spectral slopes. The transition from near-complete coherence at $\sigma < N_l$ to insignificant near-zero coherence over 1 to 2 orders of frequency range was previously found to describe the transition from 2D stratified turbulence to full 3D isotropic turbulence associated with internal wave breaking above a deep-ocean seamount (van Haren *et al*., 2016).

As indicated in the cartoon of Fig. 7, shallow relatively clear waters such as in lagoons can thus be quite different in their daily heating/cooling pattern compared with the near-surface open-ocean. During the warming phase, shear-convective turbulent overturning is generated near the seafloor due to the heating from below in addition to stable stratification from the surface down (which however is also mixed downward under windy conditions, like in the open-ocean, see, e.g., Price *et al*., 1986). During the cooling phase, the site is not only governed by free turbulent convection as is observed in the ocean, but it is seen to develop a weak stratification that supports (non)linear internal waves. This is probably because of the sudden absence of bottom heating when sun radiation and conduction stops. The associated strong reduction of near-bottom turbulence transfers upward, like in the nighttime atmosphere (Nieuwstadt, and Brost, 1986).

The observed physics processes of near-bottom heating during the warming phase and stratification during the cooling phase should be taken into account for lagoon modelling, some of which thus far adopted open-ocean warming/cooling processes (e.g., McCabe *et al*., 2010). Very shallow <0.1 m water depth lake modelling proved the possibility of near-bottom heating (de la Fuente, 2014). Future lagoon modelling should also account for the additional biophysics process of microenvironment heating of particular (dark) shallow-water corals and reef formation (Fabricius, 2006; Jimenez *et al*., 2008) that adds to the presently observed physics process of heating from below in a shallow lagoon.

The present observations were from above a depression in a more or less white bottom. The same process with similar temperature differences was seen in observations from above the darker bottom just outside Apooiti marina (Tahiti) using the same instrumentation (not shown) and episodically in shallow water of a tidal flat sea on a calm summer day (van Haren, 2019). The physics process of solar pond heating in a stable salt stratification (von Kalecsinksy, 1902) may be observable in open-sea connected and -flushed lagoons and even in the open ocean after rainfall, as may have been the case (but not discussed) in observations by (Soloviev, and Lukas, 1997). However, near-bottom heating has not been observed in turbid waters like near a North Sea beach on a quiescent summer day with no wind-wave action (van Haren *et al*., 2012), which must be due to the lack of solar radiation penetration at the bottom in waters having a few 0.1 m of visibility.

## ACKNOWLEDGMENTS

I thank C. Millot for his critical comments and the overboard operations from the S/V "ORUA", M. Laan for his T-sensor assistance and F. Bosveld for his explanation on the contrast between water- and air-convection. NIOZ temperature sensors have been partially funded by NWO, the Netherlands Organization for the Advancement of Science.

**Figure Captions**

**Fig. 1**. Mooring and site. (a) Mooring line laid out on the vessel's anchor chain guide. For reference, the high-resolution temperature T-sensor housings are 0.18 m long. The two bottles constituting the top of the mooring provide 0.5 kg net buoyancy. The uppermost T-sensor tip is 0.05 m below the bottle-necks. The low-resolution pressure-temperature-salinity PTS-sensor is between uppermost-first and second T-sensors. Originally the distance between the T-sensors was 0.38 m (except 0.45 m between the upper two), but by tying knots it was reduced to about 0.35 m. The lowest sensor is about 0.15 m above the 1 kg lead weight. (b) Mooring site (circle) in the lagoon near Bora Bora atoll in the south-Pacific.

**Fig. 2**. Entire time series of Bora Bora deployment in August 2016. Time is given in yeardays local solar time 'LST', computed from the mooring site longitude. The yellow bars indicate periods of daytime, between sunrise and sunset. (a) Temperature for all sensors in the color range [blue, red, magenta, green, cyan] from upper (first) to lower (fifth) sensors. (b) Temperature difference relative to the lowest sensor #5 (using the same colors as in a., so that the zero-line is cyan). (c) Low-pass filtered (~40 cpd, cycles per day cut-off) temperature difference between three sets of neighbouring sensors.

**Fig. 3**. Five day representative detail of Fig. 2. (a) Temperature, with colours as Fig. 2a. (b) Depth (pressure) variations as measured from the PTS-sensor. (c) Salinity from the PTS-sensor. The vertical scale is the same as that of a. in terms of density contributions using the local thermal expansion coefficient. (d) Temperature relative to the lowest sensor, as Fig. 2c.

**Fig. 4**. Smooth spectra for data from days 216-228. (a) Temperature variance spectra for three levels of warming phase data during daytime between xx.37 and xx.77, xx denoting an arbitrary yearday. Some spectral slopes (log-log scale) are given by straight black lines.

These lines are the same in all following variance spectra and may be used for reference. (b) As a., but for cooling phase data between day.87 and (day+1).27. (c) Coherence spectra for pairs of neighbouring sensors for warming phase data. The horizontal dashed line indicates the approximate 95% significance coherence level. (d) As c., but for cooling phase.

**Fig. 5**. As Fig. 4, but for calm days 214, 215 with turbid near-surface waters.

**Fig. 6**. As Fig. 5, but for windy days 212, 213 with clear waters.

**Fig. 7**. Sketches of vertical temperature profiles following ocean-atmosphere interactions for warming phase (left; dominant incoming solar radiation with possible wind shear stress) and cooling phase (right; dominant outgoing radiation generating convective turbulence). (a) General consensus for open-ocean (near-surface) conditions becoming stratified during warming phase and convectively well-mixed during cooling phase (e.g., Brainerd and Gregg, 1993). (b) Present observations leading to the hypothesis for shallow lagoons becoming heated from below including turbulent motions during warming phase and being weakly stratified including internal waves in the water column during cooling phase.

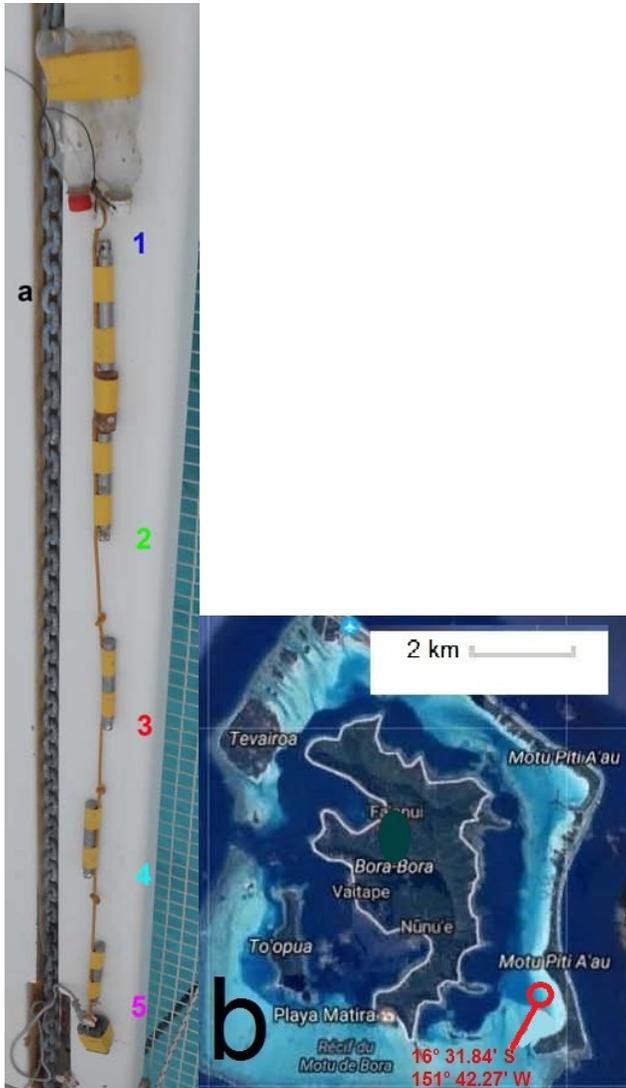

**Fig. 1**. Mooring and site. (a) Mooring line laid out on the vessel's anchor chain guide. For reference, the high-resolution temperature T-sensor housings are 0.18 m long. The two bottles constituting the top of the mooring provide 0.5 kg net buoyancy. The uppermost T-sensor tip is 0.05 m below the bottle-necks. The low-resolution pressure-temperature-salinity PTS-sensor is between uppermost-first and second T-sensors. Originally the distance between the T-sensors was 0.38 m (except 0.45 m between the upper two), but by tying knots it was reduced to about 0.35 m. The lowest sensor is about 0.15 m above the 1 kg lead weight. (b) Mooring site (circle) in the lagoon near Bora Bora atoll in the south-Pacific.

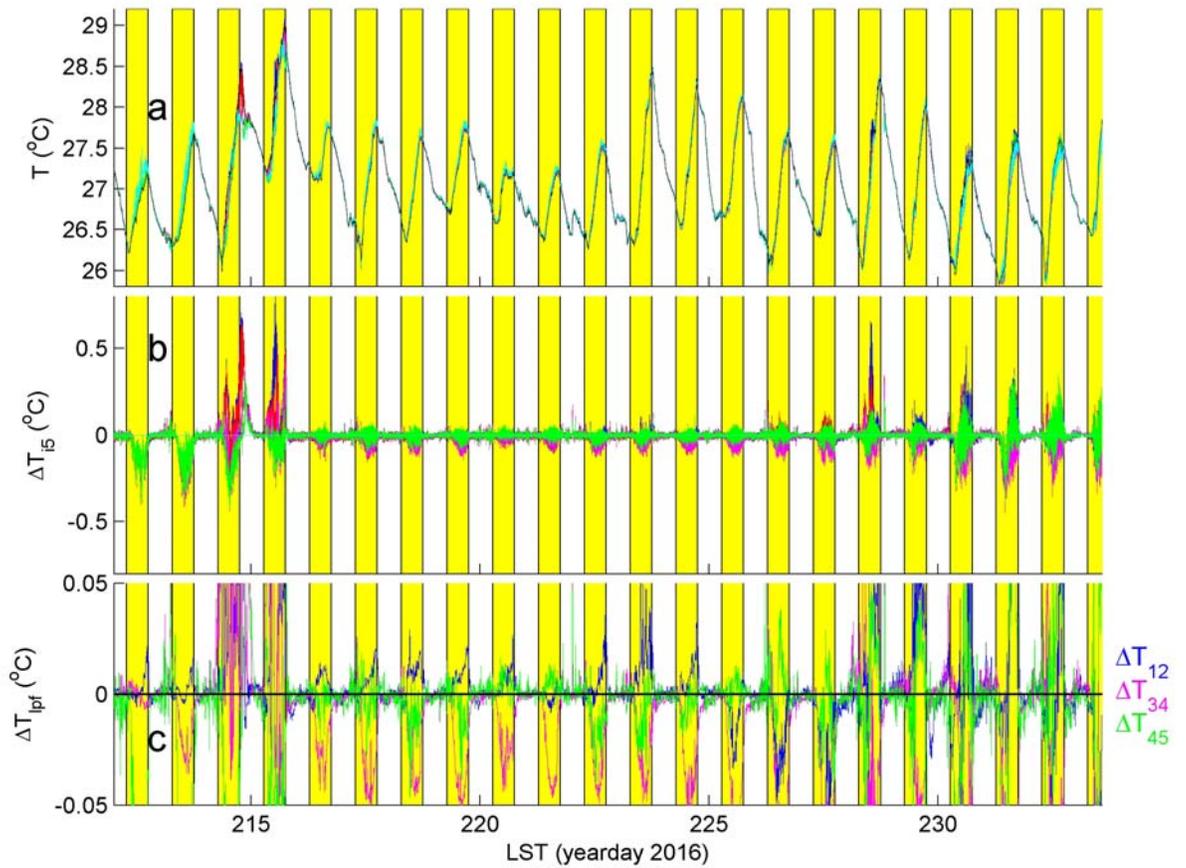

**Fig. 2**. Entire time series of Bora Bora deployment in August 2016. Time is given in yeardays local solar time 'LST', computed from the mooring site longitude. The yellow bars indicate periods of daytime, between sunrise and sunset. (a) Temperature for all sensors in the color range [blue, red, magenta, green, cyan] from upper (first) to lower (fifth) sensors. (b) Temperature difference relative to the lowest sensor #5 (using the same colors as in a., so that the zero-line is cyan). (c) Low-pass filtered (~40 cpd, cycles per day cut-off) temperature difference between three sets of neighbouring sensors.

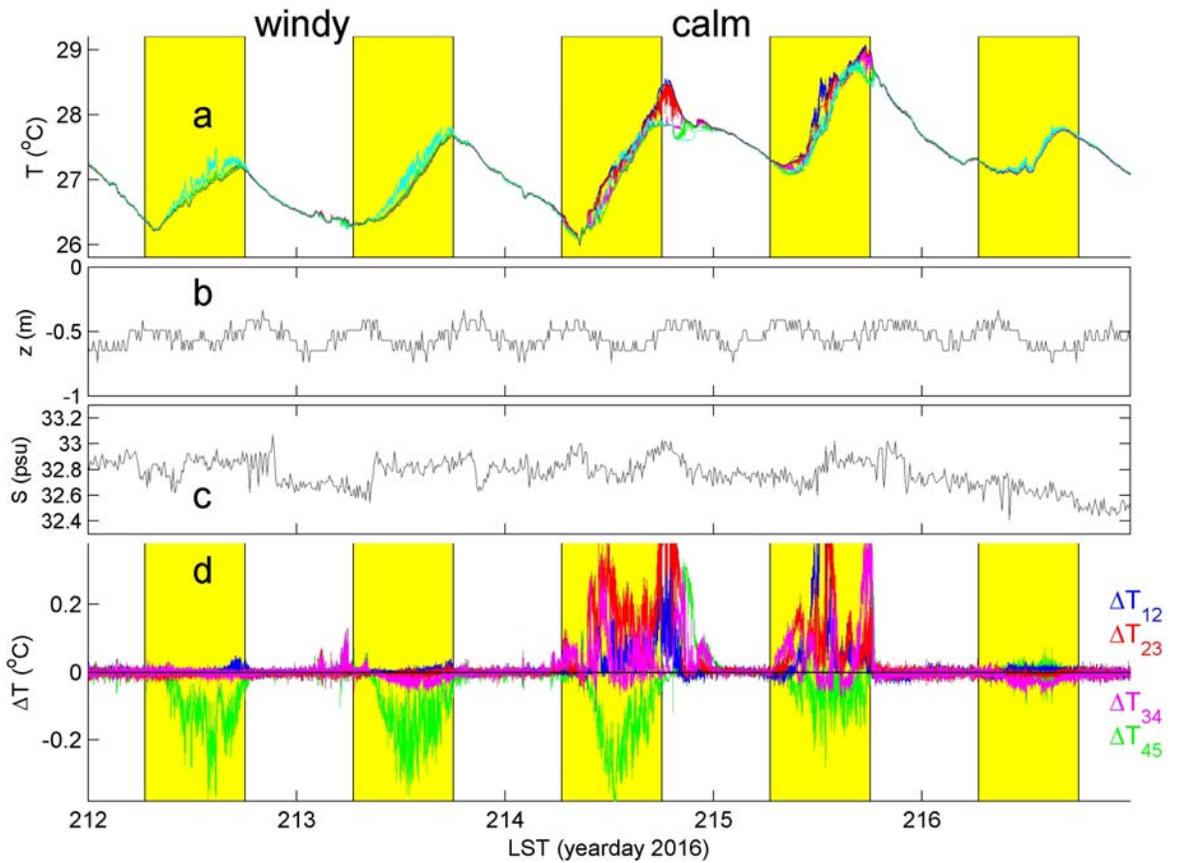

**Fig. 3**. Five day representative detail of Fig. 2. (a) Temperature, with colours as Fig. 2a. (b) Depth (pressure) variations as measured from the PTS-sensor. (c) Salinity from the PTS-sensor. The vertical scale is the same as that of a. in terms of density contributions using the local thermal expansion coefficient. (d) Temperature relative to the lowest sensor, as Fig. 2c.

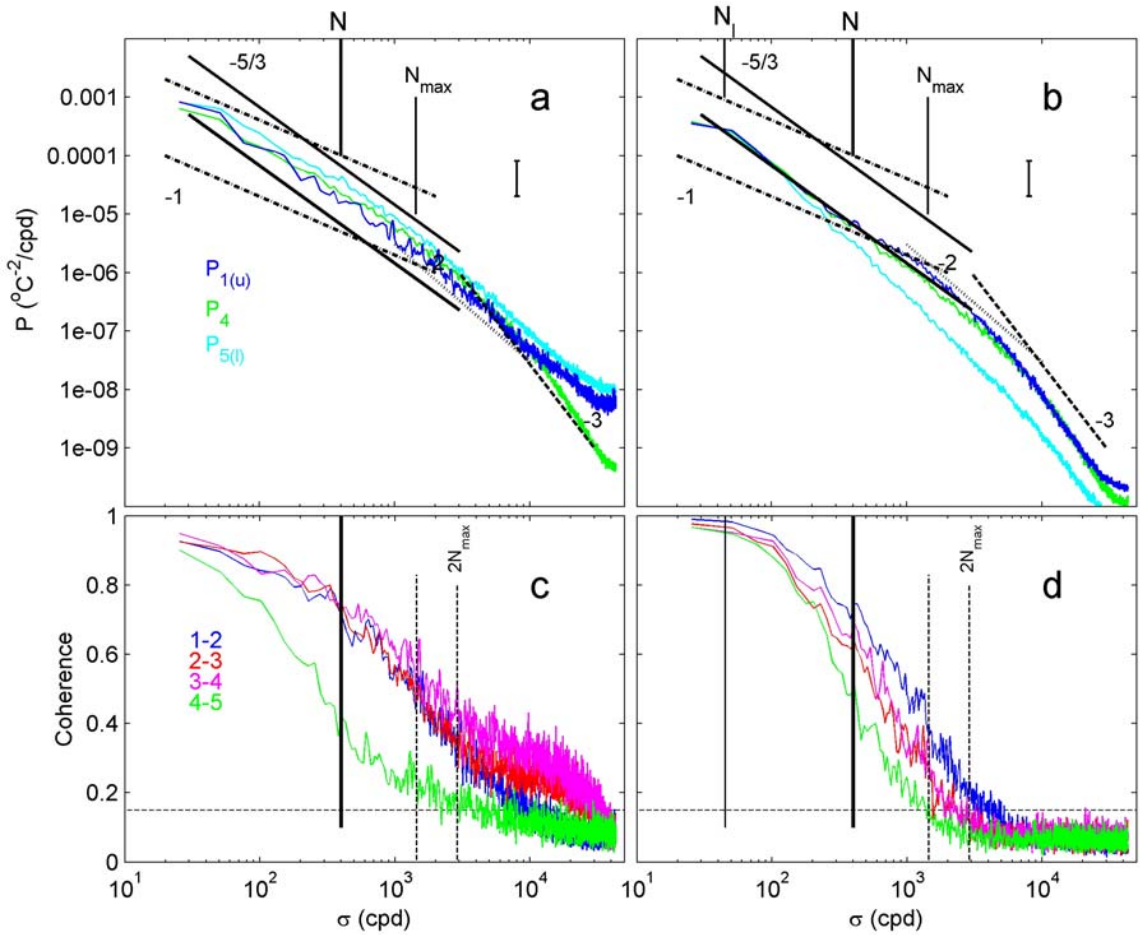

**Fig. 4**. Smooth spectra for data from days 216-228. (a) Temperature variance spectra for three levels of warming phase data during daytime between xx.37 and xx.77, xx denoting an arbitrary yearday. Some spectral slopes (log-log scale) are given by straight black lines. These lines are the same in all following variance spectra and may be used for reference. (b) As a., but for cooling phase data between day.87 and (day+1).27. (c) Coherence spectra for pairs of neighbouring sensors for warming phase data. The horizontal dashed line indicates the approximate 95% significance coherence level. (d) As c., but for cooling phase.

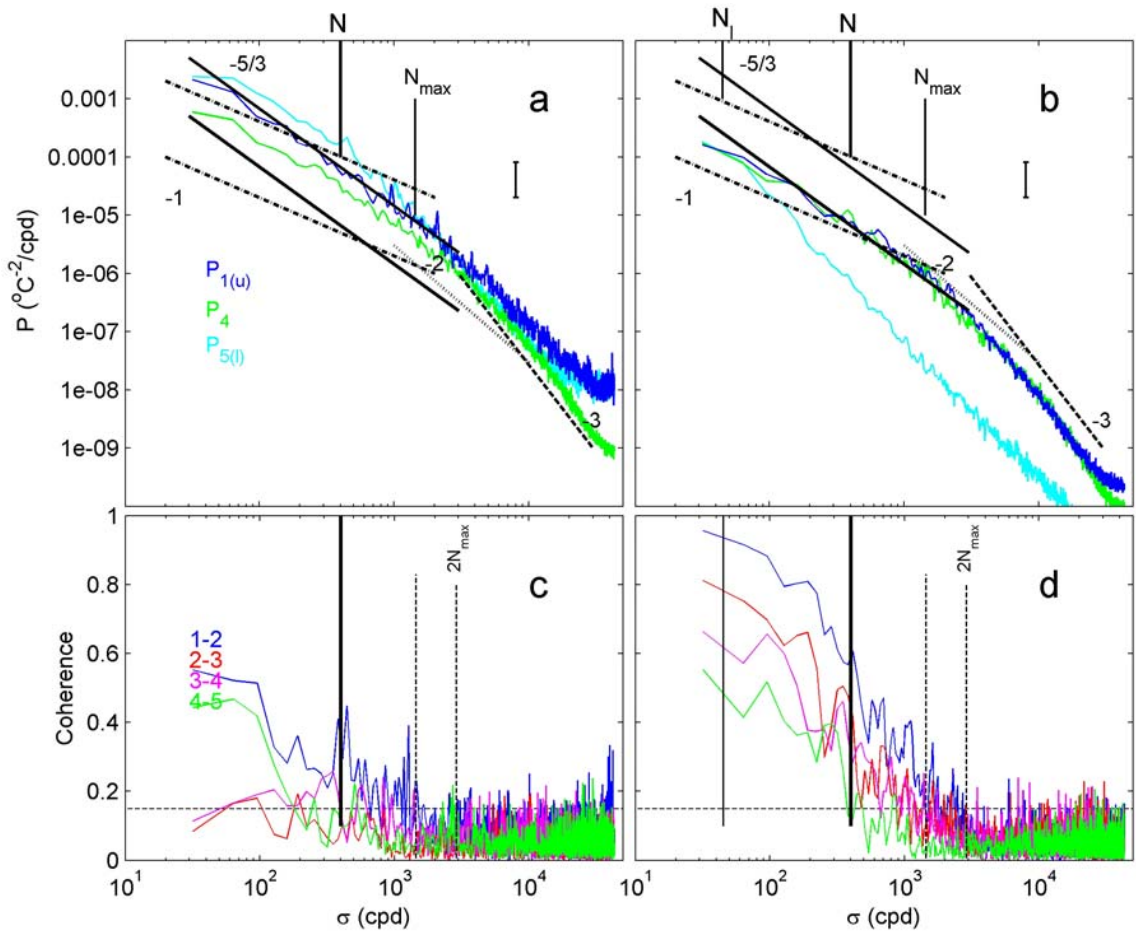

**Fig. 5**. As Fig. 4, but for calm days 214, 215 with turbid near-surface waters.

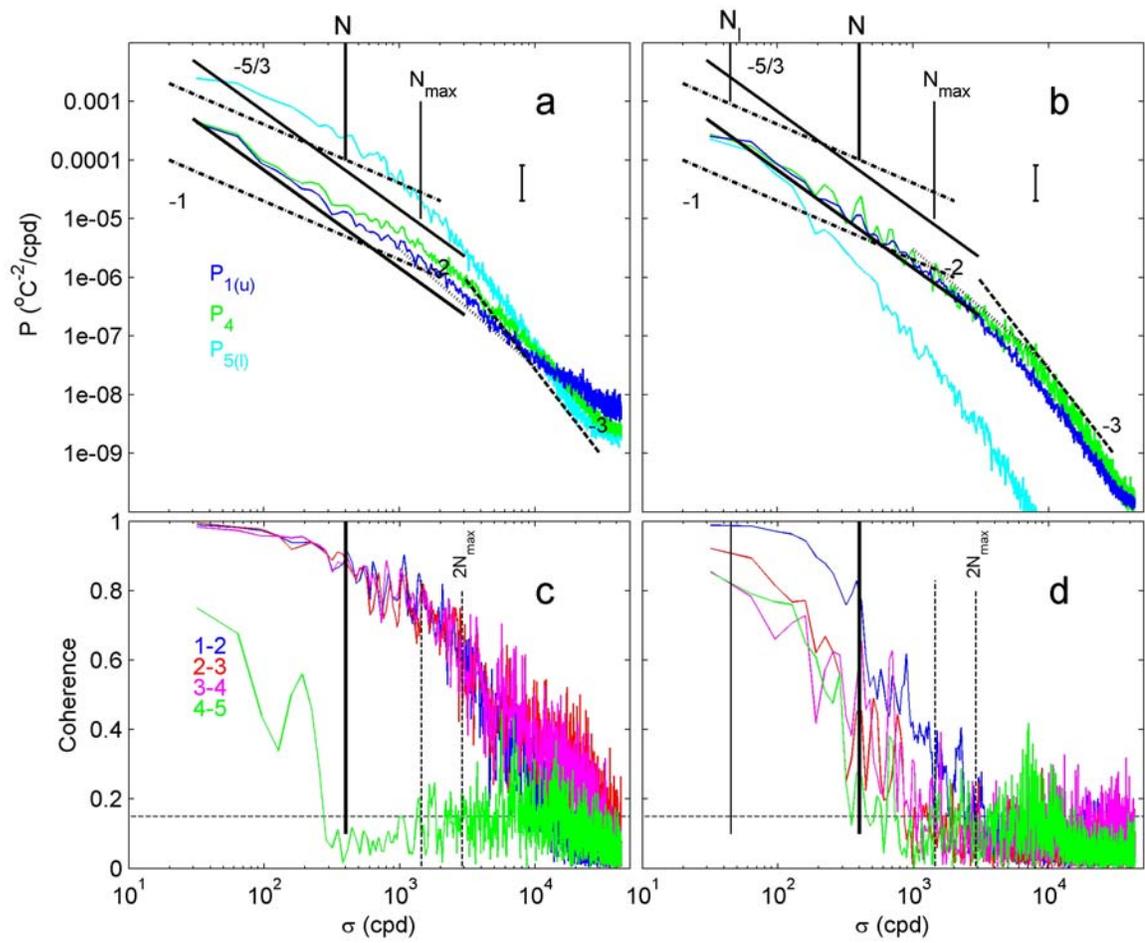

**Fig. 6**. As Fig. 5, but for windy days 212, 213 with clear waters.

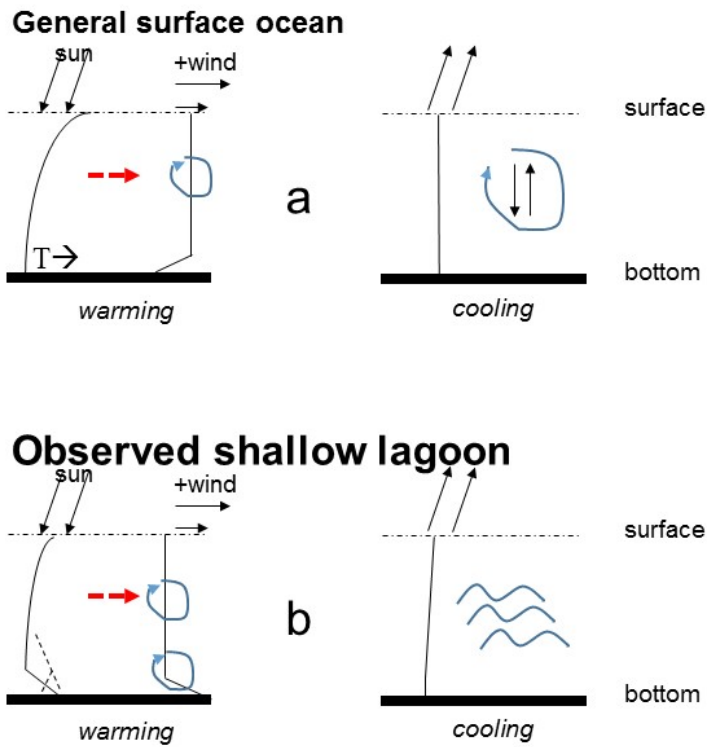

**Fig. 7**. Sketches of vertical temperature profiles following ocean-atmosphere interactions for warming phase (left; dominant incoming solar radiation with possible wind shear stress) and cooling phase (right; dominant outgoing radiation generating convective turbulence). (a) General consensus for open-ocean (near-surface) conditions becoming stratified during warming phase and convectively well-mixed during cooling phase (e.g., Brainerd and Gregg, 1993). (b) Present observations leading to the hypothesis for shallow lagoons becoming heated from below including turbulent motions during warming phase and being weakly stratified including internal waves in the water column during cooling phase.